\documentclass{aastex631}
\usepackage{amsmath,amssymb}
\usepackage{pifont}

\begin{document}

\title{Wave generation and energetic electron scattering in solar flares}

\author[0000-0002-0786-7307]{Hanqing Ma}
\affiliation{Department of Physics, University of Maryland, College Park, MD 20740, USA}

\author[0000-0002-9150-1841]{J. F. Drake}
\affiliation{Department of Physics, University of Maryland, College Park, MD 20740, USA}

\author[0000-0002-5435-3544]{M. Swisdak}
\affiliation{IREAP,
University of Maryland,
College Park, MD 20742, USA}

\begin{abstract}
We conduct two-dimensional particle-in-cell simulations to investigate
the scattering of electron heat flux by self-generated oblique electromagnetic waves. The
heat flux is modeled as a bi-kappa distribution with a $T_{\parallel}>T_{\perp}$ temperature
anisotropy maintained by
continuous injection at the boundaries. The anisotropic distribution
excites oblique whistler waves and filamentary-like Weibel instabilities.  Electron velocity distributions taken after the system has reached a steady state show that these instabilities
inhibit the heat flux and drive the total distributions towards
isotropy. Electron trajectories in velocity space show a circular-like
diffusion along constant energy surfaces in the wave frame. The key parameter controlling the scattering rate is
the drift speed of the heat flux $v_d$ compared with the electron Alfv\'en speed $v_{Ae}$, with higher drift speeds producing stronger fluctuations and a more
significant reduction of the heat flux. Reducing the density of the
electrons carrying the heat flux by 50\% does not significantly affect
the scattering rate. A scaling law for the electron scattering rate versus $v_d/v_{Ae}$ is deduced from the simulations. The implications of these results for understanding 
energetic electron transport during solar flare energy release are discussed.
\end{abstract}

\keywords{Active solar corona (1988), Plasma physics (2089), Solar flares (1496), Solar coronal waves (1995)}
% ---------------------------------------------------------------------------
\section{Introduction} \label{sec:intro}
% ---------------------------------------------------------------------------

In solar flares electrons can be accelerated to energies above 100 keV \citep{Lin2003,Krucker2007,Krucker2010}, three orders of magnitude higher than typical ambient coronal values. Because the X-rays and microwaves produced by solar energetic electrons are observed in both the corona (the location of the initial energization) and the chromosphere, transport, a crucial process that remains poorly understood, must be a key component of the dynamics of flare energy release. Observations point to a mechanism (or mechanisms) that scatter and inhibit  electron transport \citep{simoes2013implications,fleishman2022solar}.  Transport effects can also modify the energy distribution of the propagating electrons and the observed X-ray spectra, which complicates the identification of acceleration mechanisms. 

Magnetic reconnection is thought to control the release rate of magnetic energy in flares \citep{yamada2010magnetic,hesse2020magnetic}, occurring on lengthscales reaching ${10}^4$ km over timescales on the order of tens of seconds. Thus, in the absence of scattering, a relativistic electron will take less than 0.1s to escape from a flare region. However, observations of above-the looptop sources reveal that the lifetime of energetic electrons is two orders of magnitude longer than the free-streaming transit time   \citep{masuda1994loop,Krucker2007,Krucker2010,fleishman2022solar}, which suggests suppression of the transport of energetic electrons.

Observations have revealed the existence of looptop hard X-ray sources during flares \citep{masuda1994loop,krucker2008hard,guo2012properties,simoes2013implications,fleishman2022solar}. These sources are situated at or distinctly above the top of the soft X-ray flaring loops. Increased detection sensitivity indicates that the above-the-looptop hard X-ray emission is likely a common feature of all flares \citep{petrosian2002loop}. Such sources require that the transport of energetic electrons be inhibited in the acceleration region so that bremsstrahlung can produce the fluxes of hard X-rays. Further, it has been shown that the fluxes of energetic electrons in the looptop are several times (a factor of 1.7 -- 8) higher than the fluxes of electrons reaching the footpoints during the flare impulsive phase \citep{simoes2013implications}, suggesting an accumulation of electrons in the looptop. Thus,  \cite{simoes2013implications} concluded that accelerated electrons must be subjected to a trapping mechanism that holds a significant fraction of them inside the coronal loops. 

A long-standing issue in solar physics \citep{moore1980thermal} is that the observed cooling times of soft X-ray sources are much longer than the predictions from models where cooling proceeds by collision-dominated \citep{spitzer2006physics} conduction. This finding gives a powerful motivation to consider the possible limitation of heat conduction by a confinement mechanism that inhibits the transport of energetic electrons. The turbulent suppression of heat conduction is one possibility. Turbulent fluctuations in the ambient magnetic field act to enhance the angular scattering rate and thus reduce the rate of escape of energetic electrons from the acceleration region \citep{bian2018heating}. But, numerical results show that to account for the observations the turbulence generated by the energy release should persist beyond the impulsive phase, which requires an extended release of magnetic energy \citep{bian2018heating}. 

The above theoretical and observational evidence has led people to consider potential mechanisms that can scatter and confine energetic electrons in the corona. One possible mechanism is the formation of a highly localized electrostatic potential drop, in the form of a double layer (DL),  that inhibits the transport of energetic electrons \citep{li2012suppression,li2013coronal,li2014dynamics}. However, \cite{li2013coronal,li2014dynamics} found that the effectiveness of confinement by a DL is proportional to the magnitude of the DL potential drop, which scales as $T_{eh}$, the hot electron temperature. Since $T_{eh}\ll m_ec^2$, the strength of the DL is not sufficient to inhibit the transport of nonthermal electrons. Another possibility is magnetic trapping, for which an effective mirror requires significant perpendicular electron velocities. The observations of strong microwave-producing gyrosynchrotron emission, which arises mainly from a trapped population of electrons spiraling in coronal magnetic loops, also requires significant perpendicular electron energy \citep{gary2018microwave}. However, recent studies of reconnection \citep{drake2006electron,dahlin2014mechanisms,dahlin2016parallel} suggested that the primary energy gain during the acceleration process is parallel to the ambient magnetic field. Therefore, to account for the observations and facilitate magnetic trapping,  a mechanism must develop to scatter the energetic electron parallel motion into the perpendicular direction. 

A potential mechanism is scattering by oblique whistler waves \citep{roberg2016suppression,roberg2018suppression,roberg2018wave,roberg2019scattering}. The cold plasma dispersion relation of a whistler wave is
\begin{equation}
\omega=\frac{\left|k_\parallel\right|kd_e^2\Omega_e}{1+k^2d_e^2}\qquad \text{or} \qquad  \frac{v_\parallel}{v_{Ae}}=\frac{kd_e}{1+k^2d_e^2}
\label{dispersion}
\end{equation}
where $\omega$ is the wave frequency, $k_\parallel$ is the wave number along the direction of the uniform background magnetic field $B_0$, $v_{Ae}$ is the electron Alfv\'en speed and $d_e$ is the electron skin depth. The parallel phase speed $v_\parallel$ first increases then decreases with $kd_e$, with a maximum of $0.5 v_{Ae}$ when $kd_e=1$. Thus, the characteristic phase speed of whistler waves is $\sim v_{Ae}$.  According to the quasilinear theory of whistler wave scattering \citep{stix1992waves}, resonant particles diffuse in velocity space along curves of constant energy in the wave frame (i.e., the reference frame that moves with the wave parallel phase speed $v_{ph}=\omega/k_\parallel$ along $B_0$). Thus, the diffusive flux of particles is locally tangent to  semicircles centered on the parallel phase velocity $v_{ph}$ of whistler waves given by

\begin{equation}
\left(v_\parallel-v_{ph}\right)^2+v_\bot^2=constant
\label{circle}
\end{equation}
For diffusion along this path, the particle's energy is conserved in the whistler wave frame, but not conserved in the lab frame, where the particle's kinetic energy is transferred to the wave's growth. Waves and particles can participate in resonant wave–particle interactions when they fulfill the resonance condition

\begin{equation}
\omega-k_\parallel v_\parallel-n\Omega_e=0
\label{condition}
\end{equation}
where $\mathrm{\Omega}_e=eB_0/m_ec$ is the electron cyclotron frequency, $n$ is an integer that can take on positive and negative values \citep{krall1973principles}, and $B_0$ is the guiding field strength. $n=0$ gives the Landau resonance condition while $n=1,-1$ correspond to the normal and anomalous cyclotron resonances, respectively.

Significant scattering at resonant velocities can lead to the appearance of “horn-like” structures in the electron distribution function \citep{roberg2019scattering,vo2021utilizing} in which the tail exhibits circular bumps at each resonant velocity. \cite{verscharen2019self} show that the necessary condition to have wave growth is $0<v_{ph}<U_{0s}$, with $U_{0s}$ the heat flux average bulk velocity. Only in this way can the resonant particles lose kinetic energy as they diffuse and transfer energy to the growing resonant wave. Furthermore, as the wave amplitude increases, the resonance widths get larger and can eventually overlap \citep{karimabadi1992physics,roberg2016suppression,vo2022stochastic}. The particle motion before the resonances overlap is periodic, but after the overlap it becomes diffusive. As a result, particles starting at small pitch-angles can scatter past 90 degrees if the wave amplitude is above the overlap threshold. 

Heat flux suppression by oblique whistler waves in the fast solar wind is under active study \citep{verscharen2019self,vasko2019whistler,lopez2019particle,micera2020particle}. In the solar wind, the heat flux is carried by the strahl, which appears as a field-aligned beam in the electron distribution. Observations reveal evidence that the strahl is scattered into the halo \citep{maksimovic2005radial,graham2017evolution,vstverak2009radial}, the superthermal component of the solar wind. Particle-in-cell simulations \citep{lopez2019particle,micera2020particle} and quasi-linear analysis \citep{vasko2019whistler} have shown that for typical parameters the strahl electrons are capable of generating oblique whistler waves via the first anomalous cyclotron resonance (see equation \ref{condition}), which is known as the fan instability or the instability of runaway electrons \citep{kadomtsev1968electric,parail1978kinetic}. The oblique whistler waves were shown to be able to drive pitch-angle scattering of the strahl, suppressing the heat flux and resulting in formation of the halo. Thus, the observations of the scattering of the strahl by whistler waves suggest that reconnection-driven energetic electrons should also drive and be scattered by oblique whistler waves. 

Another possibility is the Weibel (also known as the filamentation) instability \citep{weibel1959spontaneously}, which is driven by an electron thermal anisotropy and can self-generate transverse magnetic perturbations with wave vectors perpendicular to the direction of the higher temperature. In the reconnection context the instability is driven when $T_\parallel \gg T_\perp$, the wavevector ${\bf k}$ is perpendicular to the ambient magnetic field and the magnetic field perturbation is perpendicular to both ${\bf k}$ and the local magnetic field. The signature of this instability is a pattern of current filaments and transverse magnetic fields stretched along the ambient field. The Weibel instability has been intensively studied theoretically \citep{yoon1987exact, yoon2007relativistic}, numerically \citep{morse1971numerical,fonseca2003three}, and experimentally \citep{allen2012experimental,huntington2015observation} and has been considered a possible mechanism for heat transfer inhibition and thermal conduction limitation in laser-plasma experiments \citep{weihan1985thermal, levinson1992inhibition}. 
In this study, under sufficiently strong heat flux conditions, the temperature anisotropy within the kappa distribution can trigger the Weibel instability. The competition between whistler and Weibel instabilities is complicated, and the Weibel instability is typically found to be transient, eventually giving way to the development of whistler waves once the temperature anisotropy has been reduced.

In previous simulations treating high-$\beta$ systems \citep{komarov2018self,roberg2018suppression,roberg2018wave} a temperature gradient was imposed across the system to drive the heat flux so that the resulting whistler waves could be evolved to steady state. In later simulations with $\beta\sim 1$ oblique whistlers were driven by an initial anisotropic kappa distribution function so the waves grew to large amplitude but then decayed due to the fast relaxation of the initial anisotropy \citep{roberg2019scattering}. In this study, we consider a system in which the heat flux arises from electrons with a bi-kappa distribution propagating parallel to the ambient magnetic field. In order to drive the system to a steady state, electrons reaching the boundaries are injected on the opposite ends of the computational domain with new velocities that ensure a continual input of heat flux. Oblique whistlers are initially excited by the anisotropic distribution through the $n = -1$ anomalous cyclotron resonance. Weibel perturbations are also evident early in the simulations. At later times, the influence of the initial distribution decays and the heat flux injection from the boundary sources continues to drive the oblique whistlers. The waves resonate with and scatter the most energetic electrons through overlapping resonances, reducing the skewness of the total distribution function and limiting the heat flux. 

% ---------------------------------------------------------------------------
\section{Simulation method} \label{sec:simeth}
% ---------------------------------------------------------------------------
We carry out 2D simulations (in the $x-y$ plane) using the PIC code {\tt p3d} \citep{zeiler2002three}, which calculates particle trajectories using the relativistic Newton–-Lorentz equations and advances the electromagnetic fields using Maxwell’s equations. There is an initial uniform magnetic field $\mathbf{B}=B_0 \hat{\mathbf{y}}$ threading the plasma, so that $v_y$ is the initial parallel velocity and $v_x$ and $v_z$ are the initial perpendicular velocities. The initial electron distribution function has two components. The first is a bi-kappa distribution with a heat flux in the positive $v_y$ direction:

\begin{equation}
f_\kappa\left(v_\parallel,v_\bot\right)=\frac{n_0}{\left(\pi\kappa\right)^{3/2}\theta_\parallel\theta_\bot^2}\frac{\Gamma\left(\kappa+1\right)}{\Gamma\left(\kappa-\frac{1}{2}\right)}\left[1+\frac{v_\parallel^2}{\kappa\theta_\parallel^2}+\frac{v_\bot^2}{\kappa\theta_\bot^2}\right]^{-\left(\kappa+1\right)}\Theta\left(v_\parallel\right)
\label{fkappa}
\end{equation}
where $n_0$ is the electron density, $\mathrm{\Gamma}$ is the gamma function, $\kappa$ is a parameter that tunes the steepness of the nonthermal tail of the distribution, and

\begin{equation}
\theta_{n}^2=v_{th,n}^2\left(\kappa-3/2\right)/\kappa
\label{eq2}
\end{equation}
is the effective thermal speed, $v_{th,n}^2=2T_n/m_e$ is the regular thermal speed, where $T_n$ is the temperature in the $n$-direction, $m_e$ is the electron mass and $\Theta\left(v_\parallel\right)$ is the Heaviside step function.

The second component is a drifting Maxwellian distribution that represents the cold return current beam (moving against $B_0$)

\begin{equation}
f_M\left(v_\parallel,v_\bot\right)=\frac{n_0}{\pi^{3/2}v_T^3\left[1+{\rm erf}{\left(v_d/v_c\right)}\right]}\exp{\left\{-\left[\left(v_\parallel+v_d\right)^2+v_\bot^2\right]/v_c^2\right\}}\Theta\left(-v_\parallel\right)
\label{fM}
\end{equation}
where $v_c^2=2T_c/m_e$ is the cold thermal speed, $v_d$ is a drift speed that ensures initial zero current, while the error function ${\rm erf}{\left(v_d/v_c\right)}$ makes the density of hot and cold particles equal. This distribution choice is motivated by flare observations \citep{krucker2010measurements,oka2013kappa,warmuth2016constraints,aschwanden2017global,glesener2020accelerated} that suggest a power-law distribution of nonthermal electrons in the energy-release and electron acceleration region.

For parameters, we chose $\kappa$ = 4, $T_y = 20T_c = 40T_x$, $v_{th,y}/v_{Ae} = \sqrt2$, $c/v_{Ae} = 5$, $\beta_y = 8\pi n_0T_y/B_0^2$ = 2. While this is a relatively high $\beta$ for the corona, specific flare observations show a hard X-ray source $\beta$ of approximately 1 \citep{krucker2010measurements}. Recent simulation results also demonstrate that magnetic reconnection can drive the outflow to a marginal firehose state in which $\beta_\parallel-\beta_\bot\approx2$ \citep{arnold2021electron}. When the high energy tail of an anisotropic kappa distribution with $\kappa=4$ is scattered to an isotropic one, particle number conservation implies the resulting energy spectral index will be 4.5 because the isotropic distribution occupies a larger phase space ($E^{0.5}$) than the anisotropic one ($E^{-0.5}$). This value is consistent with the results of a statistical study of flare parameters, which found that the spectral index value is around 4 based on a single power law fit \citep{warmuth2016constraints}. Additionally, the large value of $T_\parallel/T_\bot$ is motivated by 3D PIC simulations of reconnection showing $P_\parallel/P_\bot$ approximately equal to 100 \citep{dahlin2017role}. 

To determine the nonthermal $\beta$, we separate the kappa distribution into a combination of a Maxwellian component and a nonthermal component, where the temperature of the Maxwellian component $\left(T_M\right)$ is related to the kappa distribution temperature $\left(T_\kappa\right)$ by the equation $T_M=\left(\kappa-1.5\right)T_\kappa/\kappa$ \citep{oka2013kappa}. \cite{oka2013kappa} define the number density of the Maxwellian component $\left(N_M\right)$ by requiring it to match the differential flux of the kappa distribution at energy $T_M$. The result is

\begin{equation}
N_M=n_0\exp{\left(1\right)}\left(\frac{1}{\kappa}\right)^\frac{3}{2}\frac{\Gamma\left(\kappa+1\right)}{\Gamma\left(\kappa-\frac{1}{2}\right)}\left[1+\frac{1}{\kappa}\right]^{-\left(\kappa+1\right)}
\label{NMtoNk}
\end{equation}
The density of the nonthermal component for the parameters under consideration can be calculated as $N_{NT}\approx0.2n_0$. The nonthermal energy density is calculated with the energy relation $n_0 T_\kappa=N_MT_M+N_{NT}T_{NT}$ with $T_\kappa \approx T_\parallel/3$. The resulting nonthermal beta is 0.17, which is slightly larger than the observed value of around 0.1 near the flare looptop \citep{fleishman2020decay}, but quite close to the 0.16 of another observation \citep{emslie2004energy}. 

The simulation domain lengths are $L_y = 163.84\ d_e$ and $L_x = L_y/4$, where $d_e = c/\omega_{pe}$ is the electron skin depth, and $\omega_{pe}={(4\pi n_0e^2/m_e)}^{1/2}$ is the electron plasma frequency. The simulation time step is $dt = 0.008\Omega_e^{-1}$, where $\Omega_e = eB_0/m_ec$ is the electron cyclotron frequency. Ions are initialized with a Maxwellian distribution of temperature $T_i = T_c$, and do not play an important role due to the large mass ratio $m_i/m_e$ = 1600. The simulation uses 560 particles per species per cell and has a grid of 1024 by 4096 cells.

We set a boundary source to continuously inject heat flux into the system in order to drive the system to a steady state.  Due to the initial distribution function (the total of equations (\ref{fkappa}) and (\ref{fM})), the heat flux moves upwards in the box while a cold return current moves downwards. Electrons that encounter the top boundary are reinjected at the bottom of the box with random positions and new velocities that satisfy the kappa flux distribution

\begin{equation}
f_{\kappa f}\left(v_\parallel,v_\bot\right)=\frac{2\left(\kappa-1\right)}{\pi\kappa\left(\theta_\parallel\theta_\bot\right)^2}v_\parallel\left[1+\frac{v_\parallel^2}{\kappa\theta_\parallel^2}+\frac{v_\bot^2}{\kappa\theta_\bot^2}\right]^{-\left(\kappa+1\right)}\Theta\left(v_\parallel\right)
\label{fkappaflux}
\end{equation}
where $\theta_{n}^2$ and other parameters are the same as the initial kappa distribution, Eq.~(\ref{fkappa}). The kappa flux distribution models particles escaping from a source. Similarly, when electrons hit the bottom boundary, they are re-injected at the top of the box with random positions and new velocities that satisfy the drift Maxwellian distribution with identical parameters as in the initial distribution. The boundary condition in the $x$ direction is periodic and ions, which are essentially only a charge-neutralizing background, simply follow periodic boundary conditions in all directions.  The electromagnetic boundary conditions are set to be periodic in both the parallel and perpendicular directions.  The ratio $v_d/v_{Ae}$, where $v_d$ is the mean parallel drift speed of the heat flux (see equations \ref{fkappa} and \ref{fkappaflux}), is expected to be a key parameter controlling the degree to which the electron distribution is in resonance with the excited waves.  In the simulations presented here we vary this parameter by changing the density $n_0$.

%-----------------------------------------------------------------------------
\section{Simulation Results}
%-----------------------------------------------------------------------------

\begin{figure}[ht!]
\centering
\includegraphics[scale=0.8]{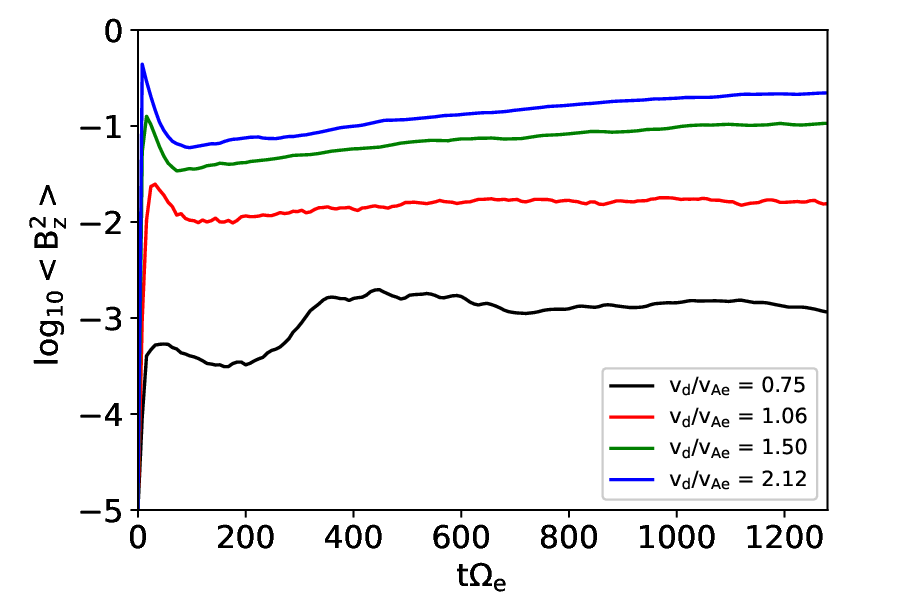}
\caption{The evolution of the box-averaged out-of-plane magnetic field fluctuations for different cases of $v_d/v_{Ae}$, where $v_d$ is the mean parallel drift speed of the heat flux and $v_{Ae}$ is the electron Alfv\'en speed.} 
\label{Fig.1}
\end{figure}

The initial distribution drives magnetic fluctuations unstable in the system. Figure \ref{Fig.1} shows the box-averaged magnetic field fluctuation $\langle B_z^2 \rangle$ versus time for different values of $v_d/v_{Ae}$, where $v_d$ is the mean parallel drift speed of the heat flux (Eq.~(\ref{fkappa})). The magnetic fluctuations grow rapidly, reach a peak value  and, after a short decay, again grow slowly until flattening to a steady state at late time. The magnetic fluctuations reach larger amplitude for higher drift speeds, where more electrons have a velocity greater than the whistler phase speed and are therefore able to drive the waves. 

\begin{figure}[ht!]
\centering
\includegraphics[scale=0.8]{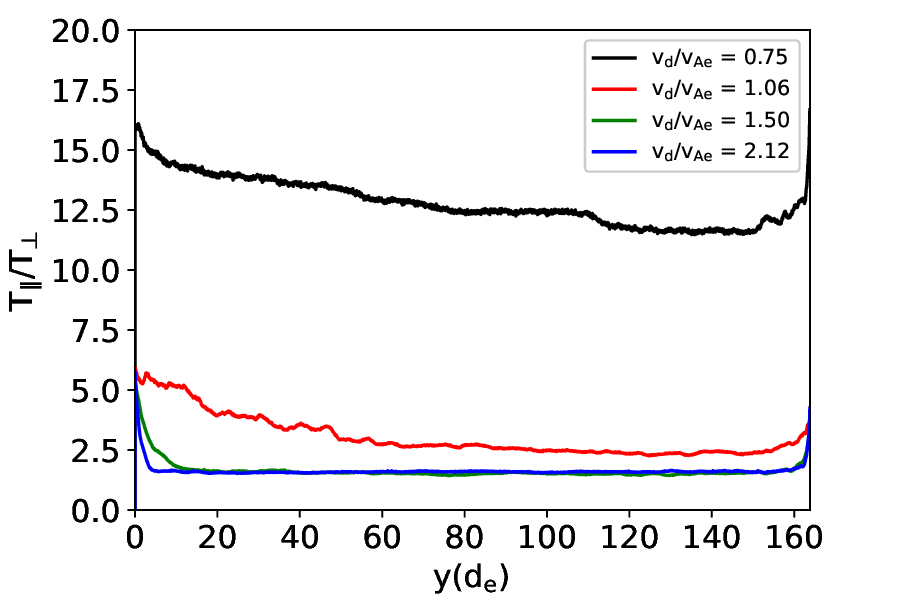}
\caption{Ratio of the parallel electron temperature $T_{eyy}$ to the perpendicular temperature $(T_{exx}+T_{ezz})/2$ of the total distribution at the end of the simulation for different values of $v_d/v_{Ae}$, where $v_d$ is the mean parallel drift speed of the heat flux and $v_{Ae}$ is the electron Alfv\'en speed.} 
\label{Fig.2}
\end{figure}

Figure \ref{Fig.2} shows the temperature anisotropy averaged over $x$ versus distance $y$ at the end of the simulations for different drift speeds. The higher drift speed cases (red, green and blue curves) exhibit roughly the same anisotropy at $y=0$ as the systems have lost all information regarding the $t=0$ initial conditions, while the slower drift speed case (black curve) is still affected by the initial distribution due to the weakness of the scattering. As a result, the anisotropy remains large. For higher drift speeds (green and blue curves), the temperature anisotropy decreases quickly with distance from the boundary and reaches a minimum value around 1.5. The final distribution is more isotropic at larger value of $y$ because the injected electrons must propagate some distance into the domain before the waves are able to isotropize them. For the slower drift speed case (black curve), the anisotropy remains large  due to the small amplitude of the waves, resulting in a correspondingly weak level of scattering.

\begin{figure}[ht!]
\centering
\includegraphics[scale=0.7]{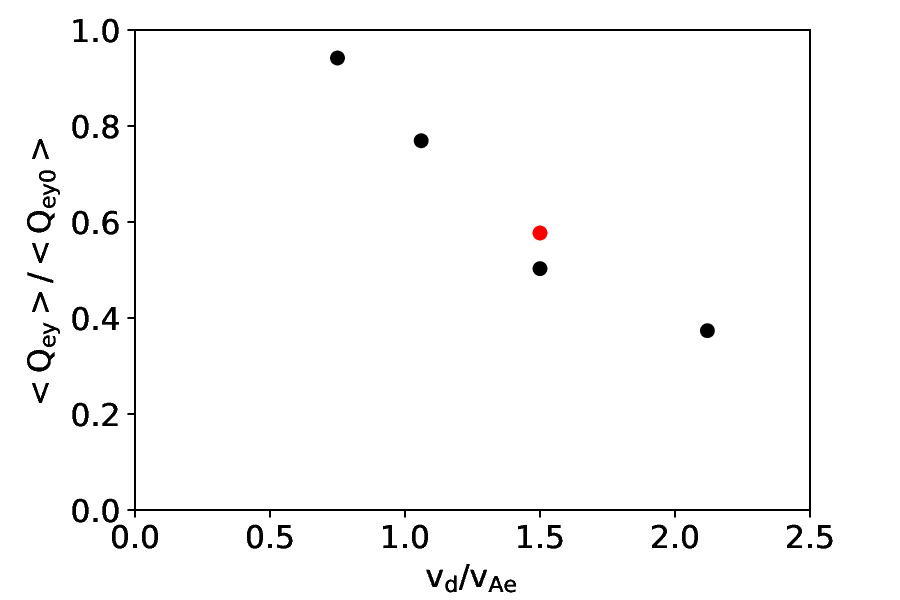}
\caption{Ratio of final to initial $y$-directed average heat flux for different values of $v_d/v_{Ae}$, where $v_d$ is the mean parallel drift speed of the heat flux and $v_{Ae}$ is the electron Alfv\'en speed. The red point corresponds to a case in which 25\% of electrons, rather than 50\%, are found in the kappa distribution (Eq.~(\ref{fkappa})) at $t=0$, while the remaining 75\% are in the drifting Maxwellian distribution (Eq.~(\ref{fM})).} 
\label{Fig.3}
\end{figure}

Shown in Figure \ref{Fig.3} is the ratio of the final to initial average $y$-directed heat flux for different drift speeds. The red point corresponds to a case where 25\% of electrons, rather than 50\%, initially satisfy the kappa distribution (Eq.~(\ref{fkappa})), while the remaining 75\% are from the drifting Maxwellian distribution (Eq.~\ref{fM})). In this case, all other distribution function parameters remain unaltered, with the exception of a reduction in the drift speed (see  Eq.~\ref{fM}) that is necessary for current balance. In this case the density of nonthermal electrons carrying the heat flux is significantly reduced  \citep{oka2013kappa,simoes2013implications,warmuth2016constraints}. The heat flux at steady state decreases with increasing drift speed, reaching a minimum value around 40\% of the initial heat flux for the highest drift speed. Notably, the reduced kappa density case (red point) does not significantly impact the reduction of the  heat flux.

\begin{figure}[ht!]
\centering
\includegraphics[scale=0.80]{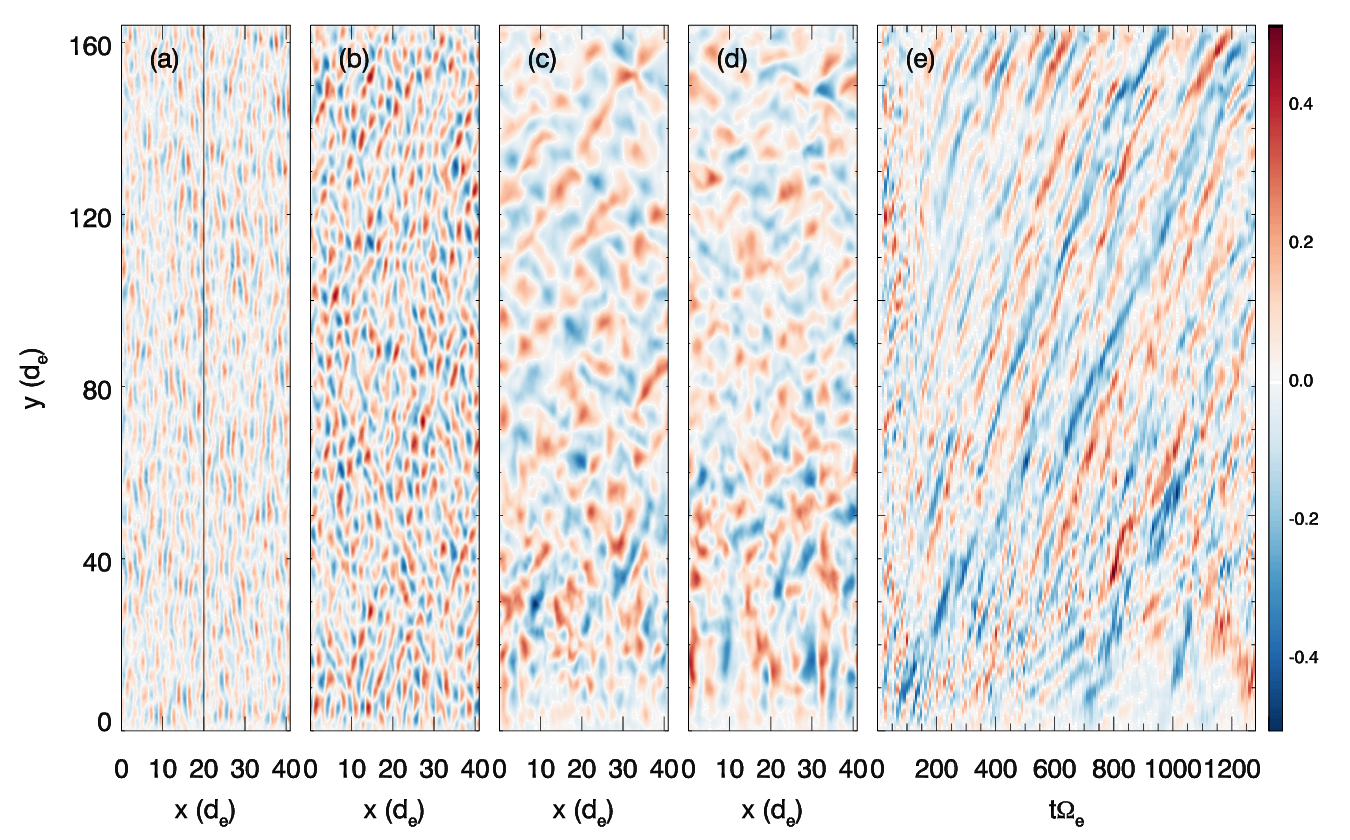}
\caption{For the case with $v_d/v_{Ae} = 1.06$ two dimensional plots of the out-of-plane magnetic field $B_z$ at (a) $t\mathrm{\Omega}_e = 15$, (b) $t\mathrm{\Omega}_e = 30$, (c) $t\mathrm{\Omega}_e = 500$. and (d) $t\mathrm{\Omega}_e = 1200$. Shown in (e) is the space-time diagram ($t-y$) of $B_z$. The black line in (a) is where the space-time plot data shown in panel (e) is taken. } 
\label{Fig.4}
\end{figure}

We now address the nature of the instabilities that drive electron scattering and specifically show that a strong Weibel instability develops in the high heat flux cases while whistlers dominate scattering at lower drift speeds. In Figure \ref{Fig.4} we show the structure of the out-of-plane magnetic fluctuations and corresponding space-time plot for $v_d/v_{Ae} = 1.06$. Shown in Fig.~\ref{Fig.4}(a) is the growth phase when the large anisotropy of the initial distribution excites the Weibel and produces filaments that fill the box along the heat flux direction. The spacing of these filaments along the $y$ direction is around $d_e$. The magnetic structure of these filaments, with the wavevector ${\bf k}$ transverse to ${\bf B}$ and the magnetic perturbation dominantly in the $z$ direction, matches the expected structure of a Weibel instability that is driven by the strong parallel heat flux.  Such Weibel-like structures are soon replaced by larger-amplitude oblique whistler waves that propagate at angles of roughly ${70}^\circ$ from the parallel direction, as shown in Figure \ref{Fig.4}(b). The whistlers are excited by the fan instability, which is driven by the $n = -1$ anomalous cyclotron resonance (see Eq.~(\ref{condition})) with the tail of the electron distribution. Figure \ref {Fig.4}(b) corresponds in time to the initial peak in the wave amplitude in Figure \ref{Fig.1}, so the initial peak and subsequent decay phase of the fluctuations should be associated with the drive from the initial distribution function. Figure \ref{Fig.4}(c) is at the time when the red curve in Figure \ref{Fig.1} begins to flatten. Oblique whistler waves with normal angles around ${45}^\circ$ are excited at the bottom of the box by heat flux injection and propagate upward. Filaments oriented along the $y$ direction, which are excited by the Weibel instability, are also observed near the lower boundary; however, they occupy only a limited region near the lower boundary of the simulation domain. 
Figure \ref{Fig.4}(d) is taken at the end of the simulation and appears similar to panel (c), showing that the system has reached a steady state. 

Shown in Fig.~\ref{Fig.4}(e) is a space-time plot of $B_z$, where the data are collected from a strip extending over the entire $y$ range of the domain at a fixed $x$ position (vertical black line in Figure \ref{Fig.4}(a)). The initial peak-and-decay phase of the magnetic field fluctuation appears as a transient period before  $t\Omega_e$ = 100, after which upward-moving whistler waves appear throughout the entire box. After around $t\Omega_e$ = 100, the bottom boundary continuously generates upward-moving whistler waves, showing that the instability from heat flux injection is responsible for the fluctuations at late time. The phase speed of the whistlers over most of the domain remains nearly constant during the entire simulation following the transient Weibel instability phase at very early time. 
%The upward wave patterns that exhibit a smaller phase speed and emerge after $t\Omega_e>400$ correspond to the quasi-parallel whistler waves that were discussed earlier.  

\begin{figure}[ht!]
\centering
\includegraphics[scale=0.65]{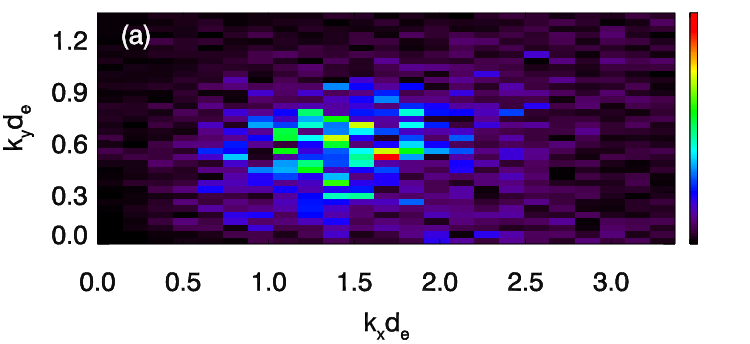}
\includegraphics[scale=0.65]{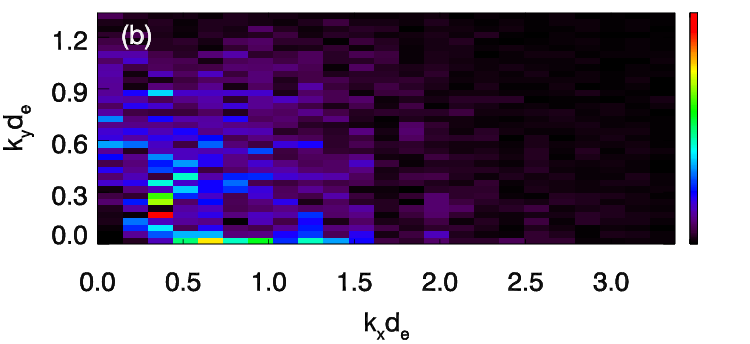}
\caption{Power spectrum of the out-of-plane magnetic field $B_z$ from the simulation in Figure \ref{Fig.4} at: (a) $t\Omega_e = 30$ (see Fig.~\ref{Fig.4}(b)); and (b) $t\Omega_e = 1200$ (see Fig.~\ref{Fig.4}(d)).}
\label{Fig.6}
\end{figure}

Figure \ref{Fig.6} shows the power spectrum of the out-of-plane magnetic field for the simulation shown in Fig.~\ref{Fig.4}. Figure \ref{Fig.6}(a) corresponds to the oblique whistler waves induced by the initial distribution (see Fig.~\ref{Fig.4}(b)), exhibits peaks around $k_xd_e\sim1.5$ and $k_yd_e\sim0.6$, and shows power extending over a broad range of wave numbers. The wave normal angle for the highest-amplitude mode is around 70$^{\circ}$, but the range of wave numbers indicates the coexistence of several oblique waves with slightly different phase speeds. The parallel phase speed $\sim 0.5 V_{Ae}$ measured in Figure \ref{Fig.4}(b) corresponds to $kd_e \sim 1$ from the dispersion relation of Eq.~(\ref{dispersion}), a value consistent with the power spectrum in Fig.~\ref{Fig.6}(a). In the steady state, there are two peaks in the power spectrum, as shown in Figure \ref{Fig.6}(b). One is the oblique mode centered around $k_xd_e\sim 0.3$ and $k_yd_e\sim 0.2$, which corresponds to the oblique whistler wave. The other is a perpendicular mode centered around $k_xd_e\sim0.6$ and $k_yd_e = 0$, which arises from the filaments appearing near the bottom boundary of Figure \ref{Fig.4}(d). The parallel phase speed $\sim 0.2V_{Ae}$ measured at the end of the simulation gives $kd_e \sim 0.1$ from the dispersion relation of Eq.~(\ref{dispersion}), which is consistent  with Figure \ref{Fig.6}(b). 

\begin{figure}[ht!]
\centering
\includegraphics[scale=0.8]{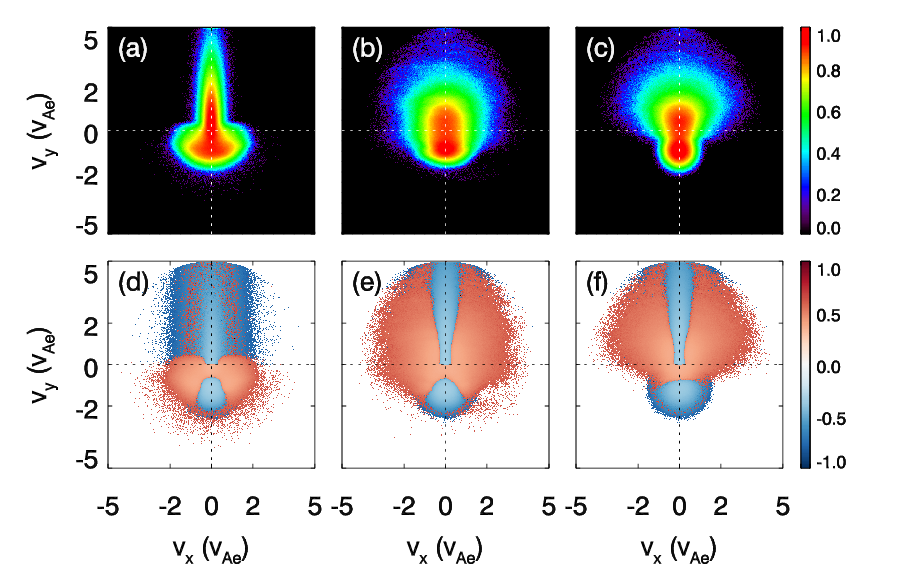}
\caption{Electron distribution functions in $v_x-v_y$ phase space at the end of the simulation of Figure \ref{Fig.4} at different $y$ positions in the box. The sampling area of each position is the entire $x$ range and 1\% of the total $y$ range. Shown in panels (a-c) are the electron velocity distribution functions on a logarithmic scale near the bottom, middle and top of the box, corresponding to positions at 0\%, 80\% and 99\% of the total $y$ range, respectively.  Panels (d-f) are comparisons to the initial distribution, $\log{\left(\left|f_{end}-f_{ini}\right|\right)}\ {\rm sgn}\left(f_{end}-f_{ini}\right)$, at the same positions, where $sgn$ is the sign function. } 
\label{Fig.7}
\end{figure}

The electron velocity-space distribution functions of Fig.~\ref{Fig.7} demonstrate the impact of the scattering of electrons by the wave activity. Shown in panels (a-c) are distributions from the end of simulation at different positions in the box. Figure \ref {Fig.7}(a), from the bottom of the simulation domain, is a kappa-like distribution in the upper half of the panel because of heat flux injection at the bottom boundary. The return current in the negative $v_y$ plane is broadened in the $v_x$ direction, due to scattering by whistler waves through the cyclotron resonance. 
%The large value of $T_\perp/T_\parallel$ for these electrons may result in a temperature anisotropy instability.
In the upper half of the box, shown in Fig.~\ref{Fig.7}(b), the total distribution has become nearly isotropic. The upward moving electrons have been strongly scattered. There is no obvious horn-like structure, which is likely a consequence of the overlap of several resonance circles \citep{vo2022stochastic} and is implied by the broad range of wave numbers shown in Figure \ref{Fig.6}. At the top of the box, panel (c), the cold current injection from the top boundary produces a beam feature in the negative $v_y$ component, while the positive $v_y$ distribution is even  more isotropic than in panel (b). Panels (d-f) show the difference between the final and initial distributions at the respective positions. Panel (d) shows that the return current has been heated, and in panels (e,f) the decrease in the parallel heat flux and the significant increase in the perpendicular velocity are direct evidence of the scattering of the nonthermal electrons by whistler waves.

\begin{figure}[ht!]
\centering
\includegraphics[scale=0.8]{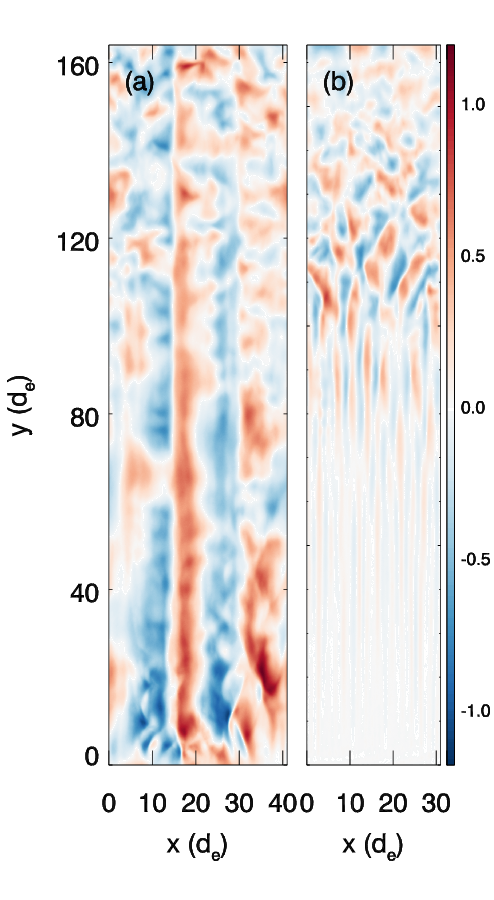}
\caption{Two dimensional plots of the out-of-plane magnetic field $B_z$ at  $t\mathrm{\Omega}_e = 1280$. In (a) the normal setup with $v_d/v_{Ae}=1.50$ (as shown in Figs.~\ref{Fig.1} and \ref{Fig.2}) and in (b) the case with the reduced number density carrying the heat flux.} 
\label{Fig.8}
\end{figure}

Shown in Fig.~\ref{Fig.8} is the structure of the out-of-plane magnetic fluctuations at the end of the simulations for the reduced number density case and the corresponding baseline run (both with $v_d/v_{Ae}=1.50$). The initial peak and decay phases are similar to Figure \ref{Fig.2} (a-b) with features such as the initial filament and subsequent oblique whistler mode, and therefore we omit those times here for simplicity. For the case with equal number densities carrying the heat flux and return current in Fig.~\ref{Fig.8}(a), the filament structure from the Weibel instability forms at the bottom of the domain and extends upward, threading the entire system. In Fig.~\ref{Fig.8}(b), where the number density of electrons carrying the heat flux is reduced, the Weibel instability is weaker and filaments only extend to the middle of the box before they are replaced by oblique whistler waves. 

\begin{figure}[ht!]
\centering
\includegraphics[scale=0.7]{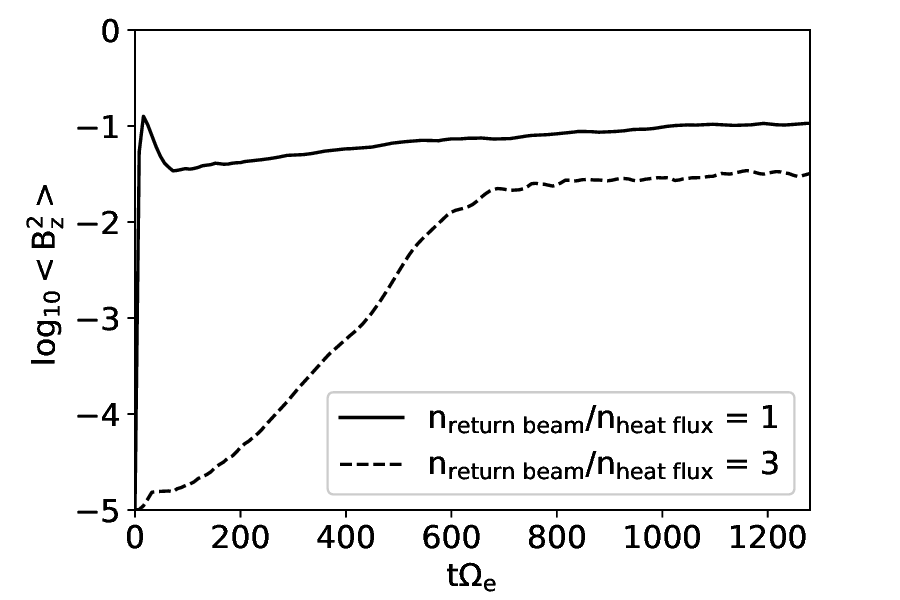}
\caption{Box-averaged out-of-plane magnetic fluctuation amplitude $B_z^2$ as a function of time for the simulation with the reduced number density of electrons carrying the heat flux (dashed curve) and the corresponding base case (solid curve). Both have $v_d/v_{Ae}=1.50$. } 
\label{Fig.11}
\end{figure}

Shown in Fig.~\ref {Fig.11} is the time evolution of the box-averaged magnetic fluctuation energy $\langle B_z^2 \rangle$ for the base case with $v_d/v_{Ae}=1.50$ and the case with reduced number density carrying the heat flux (corresponding to the red point in Fig.~\ref{Fig.3} and Fig.~\ref{Fig.8} panel (b)). When the initial heat flux is reduced by half (with the same drift speed), the wave energy before $t\Omega_e$ = 400 is small due to the reduced population of electrons available to drive the instability. Eventually the slowly growing waves reach finite amplitude and the fluctuation level flattens around $t\Omega_e$ = 800. The late time value of the wave amplitude is only modestly reduced from that of the base case, demonstrating that strong oblique whistler waves can be excited even when the number density of the electrons carrying the heat flux is reduced.

\begin{figure}[ht!]
\centering
\includegraphics[scale=0.8]{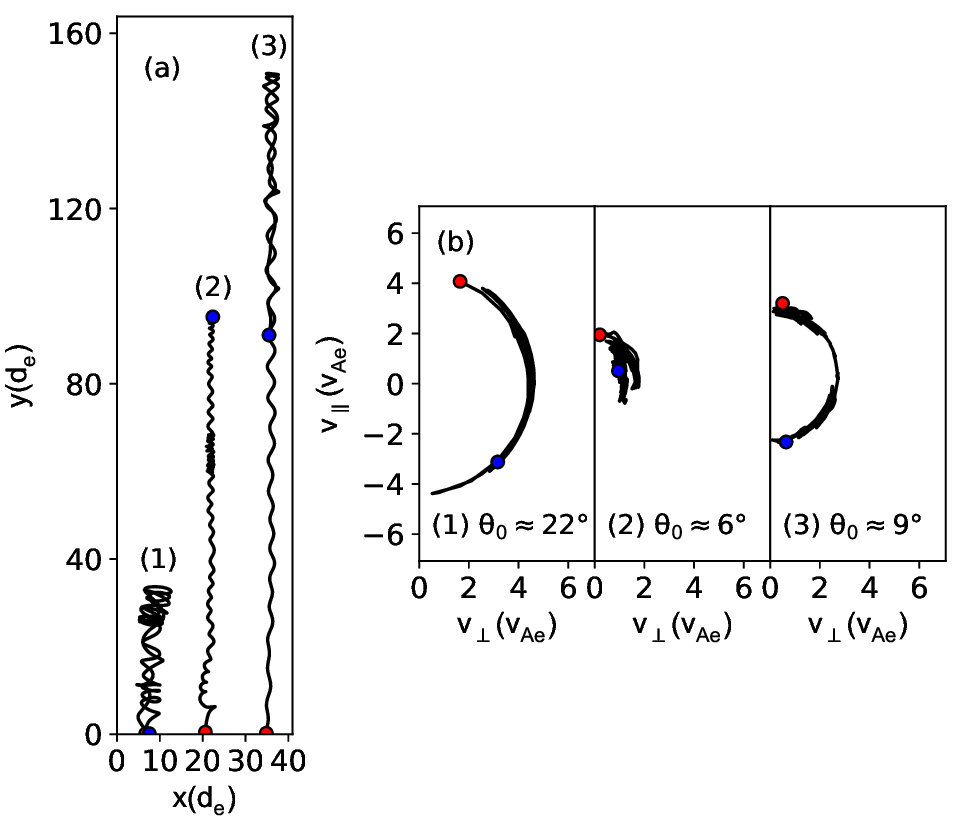}
\caption{(a) Electron trajectories in the $x-y$ plane for the strong drift case with $v_d/v_{Ae}=2.12$. Three electron trajectories are shown with red dots indicating the starting positions and blue dots the end positions. The trajectories are from the middle of the simulation (around $t\Omega_e = 900$). (b) Electron velocity trajectories correspond to the three cases in panel (a) in $v_\parallel-v_\bot$ space. Red dots indicate the starting positions and blue dots are the end of the trajectories. $\theta_0$ gives the initial pitch angle.} 
\label{Fig.9}
\end{figure}

Shown in Fig.~\ref{Fig.9} are the self-consistent trajectories of several electrons in the $x-y$ and $v_\parallel-v_\bot$ planes, extracted from the particle-in-cell simulation with $v_d/v_{Ae}=1.50$. The electrons all start from the bottom boundary with initial parallel velocities significantly greater than their perpendicular velocities and so are typical of electrons carrying the heat flux. In panel (a), the waves in the system are strong enough to completely reverse the direction of the electrons (they are scattered past $90^o$ in pitch angle). In panel (b), the trajectories of case (1) and (3) exhibit diffusion along a nearly complete semi-circle (see Eq.~(\ref{circle})) and scatter over 90 degrees in pitch angle, reversing their direction along the ambient magnetic field. The particle in case (2) also scatters past 90 degrees but reveals a more complex trajectory as it moves from a larger to a smaller radius in velocity space. The trajectories demonstrate that the reduction of electron parallel velocity is accompanied by an increase of perpendicular velocity. Thus, the scattering of whistler waves has significant consequences: it directly inhibits the heat flux while at the same time boosting the potential for magnetic trapping by increasing the pitch angle. Both effects will boost the confinement of energetic electrons in flares. 

\begin{figure}[ht!]
\centering
\includegraphics[scale=0.7]{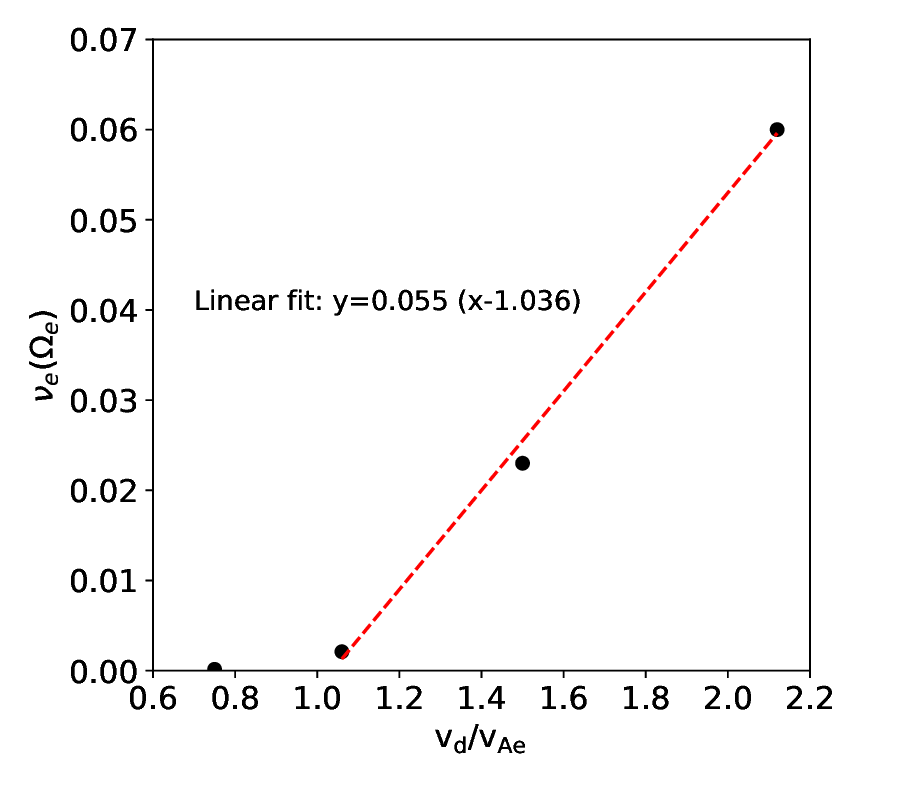}
\caption{The averaged wave-electron scattering rate at late time for the four runs in Fig.~\ref{Fig.1} as a function of $v_d/v_{Ae}$, the ratio of mean parallel drift speed of heat flux to electron Alfv\'en speed. The scattering rate is calculated from Eq.~(\ref{SSscatteringrate}) and is averaged over the ranges of $0\lesssim y \lesssim 80d_e, 0\lesssim y \lesssim 60d_e, 0\lesssim y \lesssim 10d_e,0\lesssim y \lesssim4d_e$ for the drift speeds $v_d/v_{Ae}=0.75$, $1.06$, $1.50$, and $2.12$, respectively. The red dashed line shows a linear  fit to the data above the threshold $v_d/v_{Ae}\approx 1$. }
\label{Fig.12}
\end{figure}

We now compute the wave-particle scattering rate that leads to the reduction in electron anisotropy and examine its scaling with parameters. There is no single such formalism for determining the rate because of the nonlinear nature of different scattering processes (e.g., weak scattering as described by quasilinear theory versus trapping in finite amplitude waves) \citep{gary2000electromagnetic}. For example, \cite{hamasaki1973relaxation} used quasilinear theory to determine the scattering rate of the whistler anisotropy instability while others have approximated the scattering due to the proton cyclotron instability by finding the maximum exponential decay rate of the temperature anisotropy \citep{gary2000electromagnetic,nishimura2002whistler}. Here is we use the Lorenz scattering operator, which describes electrons scattering off a stationary center, to model the relaxation of the temperature anisotropy in the simulations and to deduce the scattering rate. The Lorenz operator is give by

\begin{equation}
\left(\frac{\partial f_e}{\partial t}\right)_e=\frac{\nu_e}{2}\frac{\partial}{\partial\xi}\left(1-\xi^2\right)\frac{\partial}{\partial\xi}f_e
\label{lorenzoperator}
\end{equation}
where $\xi=\cos{\theta}$ and $\nu_e$ is the electron scattering rate.  By taking moments of this equation to determine the rate of relaxation of the temperature anisotropy of the electron velocity distribution, we obtain a formula for $\nu_e$, 

\begin{equation}
\nu_e=\frac{dR/dt}{2-R-R^2}
\label{scatteringrate}
\end{equation}
where $R=T_\parallel/T_\bot$ is the temperature anisotropy of the total electron distribution. We use this equation to model the relaxation of the temperature anisotropy near the low boundary of the simulation domain that is shown at late time in Fig.~\ref{Fig.2}. The anisotropy profile has reached a steady state so the total time derivative in Eq.~(\ref{scatteringrate}) reduces to the convective derivative
\begin{equation}
\nu_e=v_y\frac{\partial R/\partial y}{2-R-R^2},
\label{SSscatteringrate}
\end{equation}
where $v_y$ is a constant equal to the heat flux drift speed at the lower boundary. The temperature anisotropy curves in Fig.~\ref{Fig.2} were smoothed with a Gaussian filter and the calculated scattering rates were averaged over the ranges of $0\lesssim y \lesssim 80d_e, 0\lesssim y \lesssim 60d_e, 0\lesssim y \lesssim 10d_e,0\lesssim y \lesssim4d_e$, for the cases of drift speed $v_d/v_{Ae}=0.75$, $1.06$, $1.50$, and $2.12$, respectively.  In Fig.~\ref{Fig.12} we plot the wave-particle scattering rate for electron anisotropy reduction from the four simulations (Fig.~\ref{Fig.1}). For the weakest drift case $v_d/v_{Ae}=0.75$, the scattering rate is very small $\left(\sim{10}^{-4}\Omega_e\right)$, while for larger drift cases the scattering rates are significantly larger, growing to $\sim 6\times{10}^{-2}\Omega_e$. The figure reveals that the scattering rate scales as a linear function of $v_d/v_{Ae}$, the ratio of the drift speed to electron Alfv\'en speed, above a threshold given by $v_d/v_{Ae}\sim1$.

% ------------------------------------------------------------------------------------
\section{Conclusions}
% ------------------------------------------------------------------------------------

Observations and theory imply that there is suppression  of the transport of energetic electrons during solar flares.  We have explored pitch-angle scattering by oblique whistler waves and Weibel-driven disturbances as a mechanism for reducing the electron heat flux. The energetic electron heat flux escaping from a flare-like system excites strong oblique whistler waves through the anomalous cyclotron resonance. These waves pitch-angle scatter the  electrons on a rapid timescale of hundreds of electron cyclotron periods, suppressing the heat flux and increasing the perpendicular velocities of electrons. At high mean drift speeds compared with $v_{Ae}$ the nonthermal electrons can also drive the Weibel instability. The resulting scattering mechanisms can operate under the generic conditions of reconnection-driven electron energization and can reduce the field-aligned electron energy flux by up to a factor of two. Of greater significance is that this scattering process increases the energetic electrons’ perpendicular velocity so that the electrons will more effectively mirror and can be trapped in the large-scale magnetic fields of the flare energy release region. This scattering mechanism can therefore suppress the escape of electrons from the energy release regions of flares, which is required for them to reach the relativistic velocities documented in observations.

In previous simulations treating whistler waves in flares, the fluctuations damp out after scattering is complete and the heat flux from the initial electron distribution is reduced \citep{roberg2019scattering}. In contrast, in the current simulations the oblique whistler waves and Weibel perturbations are maintained in a steady-state due to the continual injection of heat flux from the boundary. This mimics the continued energy release from reconnection during flares. Such a setup represents a more physically reasonable set of initial conditions and avoids the influence of arbitrary discontinuities in the initial distribution function, giving more confidence to the interpretation of the final results. We suggest, on the basis of the present simulations, that oblique whistler waves and Weibel disturbances can grow to large amplitude during flares and scatter electrons to reduce their heat flux.

Although our model was designed to study the transport of energetic electrons in flares, the model is relevant to any $\beta\sim 1$, weakly collisional plasmas with high heat flux due to the presence of nonthermal electrons. Astrophysical applications might include accretion disc coronae \citep{bisnovatyi1976accretion,liang1977accretion,uzdensky2008statistical}. Scattering by oblique whistlers of the solar wind strahl into the halo population is currently under active investigation \citep{boldyrev2019kinetic,verscharen2019self,vasko2019whistler}.  Observations at 1AU have shown that the pitch angle widths of strahl in the presence of whistler mode waves are up to 12$^{\circ}$ larger than that in the absence of whistler waves \citep{kajdivc2016suprathermal}. Large-amplitude, oblique whistlers that would limit the electron heat flux have been detected in the solar wind \citep{breneman2010observations,cattell2020narrowband}, in the magnetosphere \citep{kellogg2011large}, and in flares \citep{huang2016two}. Simulations show that oblique whistler waves excited by the strahl component under solar wind conditions can scatter the strahl distribution into the halo and reduce the strahl-carried heat flux \citep{vo2021utilizing,lopez2019particle,micera2020particle}. Our simulations are focused on the flare system, where the electron heat flux is much greater than in the case of the solar wind strahl. Nevertheless, our model clarifies the mechanism for scattering of energetic electrons by whistlers that is also active in the strahl scattering problem. 

We note that our results may be limited by the common disadvantages of particle-in-cell simulations such as small domains and short timescales compared with those of real flares. On the other hand, the scale size of turbulence associated with whistlers and the Weibel instability is at the kinetic scale so the typical scale separation issue is less of a problem than in the exploration of electron energy gain during magnetic reconnection. The initial electron velocity distribution contains an artificially sharp gradient near $v_y = 0$, but this gradient is transient in the system’s evolution and its influence disappears by the time the system reaches a steady state. Further investigation is required to elucidate the transition mechanism from the Weibel instability to oblique whistler wave. Additionally, it is worth noting that magnetic field inhomogeneities resulting from turbulence or inherent mirror structure may affect transport suppression. The confinement of energetic electrons seems to result from a combination of magnetic trapping and wave scattering, so the exploration of particle scattering in a mirror geometry should be pursued in future studies. 

\begin{acknowledgements}
The authors were supported by NASA grants 80NSSC20K0627, 80NSSC20K1813, and 80NSSC20K1277
and NSF grant PHY2109083.
\end{acknowledgements}

\bibliographystyle{aasjournal}   
\bibliography{paperbib}

\end{document}